\documentclass[conference]{IEEEtran}
\IEEEoverridecommandlockouts
\usepackage{cite}
\usepackage{amsmath,amssymb,amsfonts}
\usepackage{algorithmic}
\usepackage{graphicx}
\usepackage{textcomp}
\usepackage{tikz}
\usepackage{xcolor}
\def\BibTeX{{\rm B\kern-.05em{\sc i\kern-.025em b}\kern-.08em
    T\kern-.1667em\lower.7ex\hbox{E}\kern-.125emX}}

\begin{document}

\title{Design and Development of a Multi-Purpose Collaborative Remote Laboratory Platform}

\author{\IEEEauthorblockN{
Sven Jacobs\IEEEauthorrefmark{1},
Timo Hardebusch\IEEEauthorrefmark{1},
Esther Franke\IEEEauthorrefmark{1},
Henning Peters\IEEEauthorrefmark{1},
Rashed Al Amin\IEEEauthorrefmark{2},
Veit Wiese\IEEEauthorrefmark{2} and
Steffen Jaschke\IEEEauthorrefmark{1}}
\IEEEauthorblockA{
\IEEEauthorrefmark{1}Computer Science Education\\
University of Siegen\\
ddi@eti.uni-siegen.de}
\IEEEauthorblockA{
\IEEEauthorrefmark{2}Institute for Embedded Systems\\
University of Siegen\\
rashed.amin@uni-siegen.de}
}

\newcommand\copyrighttext{%
  \footnotesize \textcopyright \the\year{} IEEE. Personal use of this material is permitted. Permission from IEEE must be obtained for all other uses, in any current or future media, including reprinting/republishing this material for advertising or promotional purposes, creating new collective works, for resale or redistribution to servers or lists, or reuse of any copyrighted component of this work in other works.
  }
\newcommand\copyrightnotice{%
\begin{tikzpicture}[remember picture,overlay]
\node[anchor=south,yshift=10pt] at (current page.south) {\fbox{\parbox{\dimexpr0.75\textwidth-\fboxsep-\fboxrule\relax}{\copyrighttext}}};
\end{tikzpicture}%
}
\newcommand\submittedtext{%
  \footnotesize This work has been submitted to the IEEE for possible publication. Copyright may be transferred without notice, after which this version may no longer be accessible.}

\newcommand\submittednotice{%
\begin{tikzpicture}[remember picture,overlay]
\node[anchor=south,yshift=10pt] at (current page.south) {\fbox{\parbox{\dimexpr0.65\textwidth-\fboxsep-\fboxrule\relax}{\submittedtext}}};
\end{tikzpicture}%
}
\renewcommand\fbox{\fcolorbox{red}{white}}
\setlength{\fboxrule}{2pt} % Set fbox rule width to 2pt

\maketitle
\copyrightnotice

\begin{abstract}
This work-in-progress paper presents the current development of a new collaborative remote laboratory platform.
The results are intended to serve as a foundation for future research on collaborative work in remote laboratories.
Our platform, standing out with its adaptive and collaborative capabilities, integrates a distributed web-application for streamlined management and engagement in diverse remote educational environments.
\end{abstract}


\section{Introduction}
Remote labs are crucial in education due to their cost-effectiveness and ability to simulate real-world laboratory experiences. They are also particularly beneficial in fields like engineering, where they allow safe access to physical labs for all students. In addition, students of different nationalities can work together without being physically present, such as students from European university alliances.

Therefore remote laboratories offer several benefits in modern education. Nevertheless, a coherent scientific conclusion has not yet been reached on the comparison between remote and hands-on laboratories \cite{brinson.2015}. While positive cognitive, behavioural and emotional outcomes have been observed in multiple studies, further empirical research is still needed to gain a comprehensive understanding of the learning benefits of remote laboratories in higher education across multiple disciplines \cite{post.2019}. 

Hence, in this work in progress paper, we present a multi-purpose collaborative remote laboratory platform for further research on the use of remote laboratories across multiple disciplines.

\section{Related Work}
In recent years, remote laboratories have reached a significant level of maturity. Today, they can be used in a variety of educational settings, including primary schools, higher education institutions, universities and distance learning programmes, to teach a wide range of subjects \cite{heradio.2016}.
Most existing remote labs are still confined to specific applications within lectures or exercises and cannot be easily customized by educators. For a specific use case such as Field-Programmable Gate Arrays (FPGA) remote laboratories, a wide range of different architectures are used \cite{alamin.2023}. The same can be observed in remote laboratories for control education \cite{heradio.2016a}.

\begin{figure}[!t]
    \centering
    \includegraphics[width=\columnwidth]{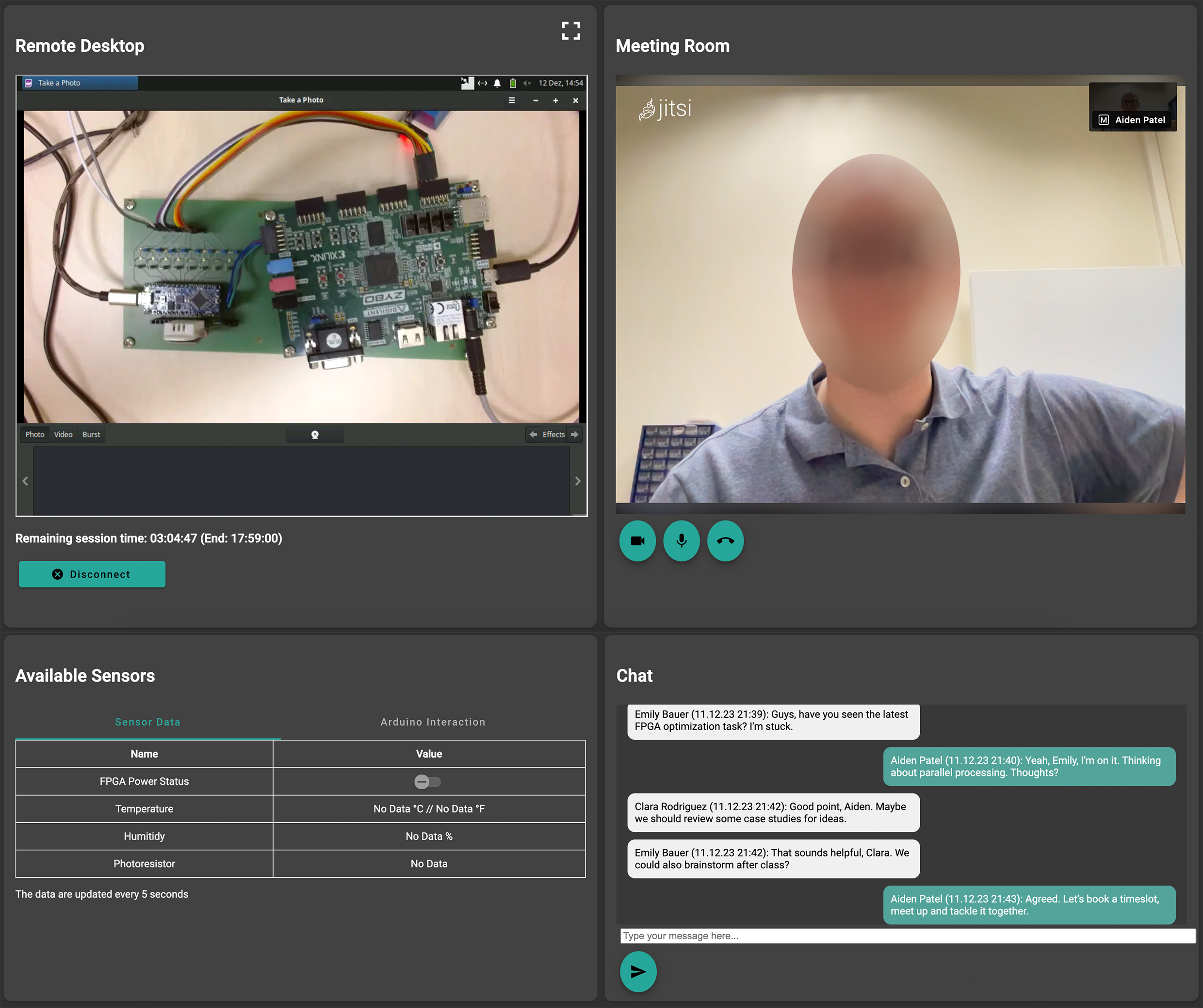}
    \caption{Collaborative workspace user interface}
    \label{fig:ui}
\end{figure}

Experimental evidence supports the usefulness of collaborative support in remote laboratories, such as increased student engagement \cite{delatorre.2013}.
Most remote laboratories still do not allow for direct collaboration beyond basic chat or video functions \cite{ashby.2008}. Our work aims to provide a highly adaptable solution for educators that enables genuine collaboration among students and teachers.

\begin{figure*}[!tbh]
\centerline{\includegraphics[width=\textwidth]{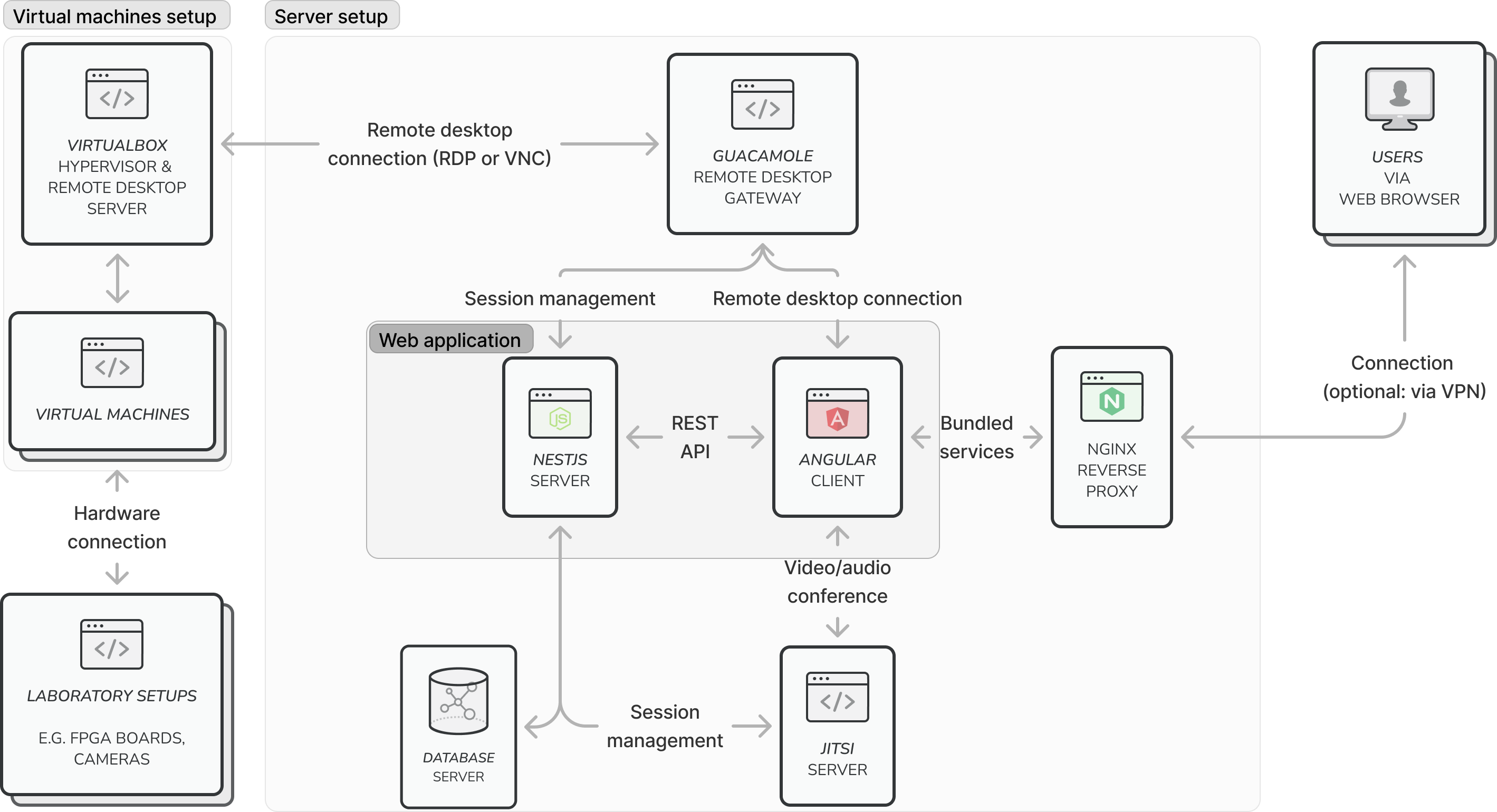}}
\caption{The implemented system for collaborative work in remote laboratories.}
\label{fig:structure}
\end{figure*}

\section{System Design \& Implementation}
The infrastructure of the implemented system is based on the key points for the design of remote laboratory architectures identified by Al Amin et al. \cite{alamin.2023}. Modifications and extensions were introduced wherever it was considered appropriate on the basis of the extended field of application in schools and universities beyond development with FPGA boards, as described below. Fig.~\ref{fig:structure} illustrates the structure of the implemented system.

\subsection{Collaborative Web-Workspace}
\begin{figure}[tb]
    \centering
    \includegraphics[width=\columnwidth]{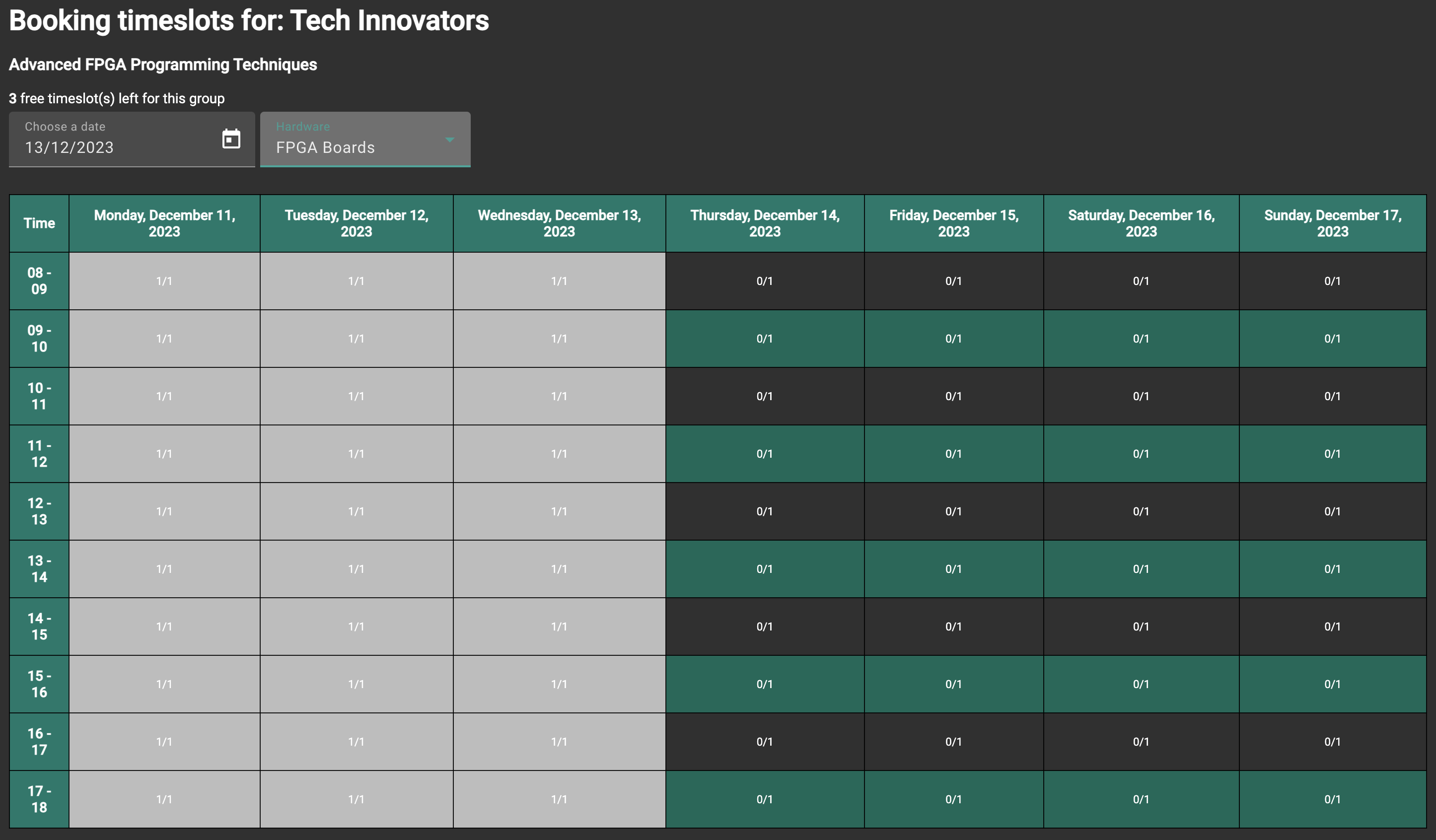}
    \caption{User interface for time slot booking}
    \label{fig:timeslotBooking}
\end{figure}

A distributed web application was developed that serves as a central hub for all components and as a single point of access for users of the system. This covers both administrative tasks (e.g. user management, configuration of laboratory setups) as well as the organization and implementation of teaching, learning and examination processes. In order to structure administrative and teaching responsibilities, users are assigned different roles with individual permissions and associated functions: administrators, teachers and students. The implemented system provides different levels of granularity for organization and collaboration within an institution: courses managed by teachers and, within those, groups of students sharing common tasks and work sessions. The latter are organized in time slots that can be self-booked by student groups for associated laboratory setups using a built-in booking tool, as shown in Fig.~\ref{fig:timeslotBooking}. To achieve a balanced laboratory utilization and a fair distribution, administrators can impose restrictions (e.g. a maximum allowed number of time slots per student group and week) on reservations.

The work sessions take place within a virtual dashboard, as illustrated in Fig.~\ref{fig:ui}. This dashboard provides users with a range of tools to facilitate successful collaboration without the need to switch between different program interfaces. A remote desktop workspace is provided where all users share a common session (screen output and input focus) on the laboratory system, comparable to multiple people sitting in front of a single computer. An integrated camera feed enables users to observe the laboratory environment in near real-time. For intra-group communication, users have access to a video and audio conference using an embedded \emph{Jitsi} instance that is automatically managed for each session. This also allows users to share their own device screen. In addition, a cross-session text chat is available for communication and note taking. An optional widget allows access to sensor data and direct control of the laboratory hardware, if applicable to the particular setup.

The designed system uses \emph{Apache Guacamole} as a gateway for remote desktop connections to the laboratory setups. This allows seamless integration into the web application, rendering a separate remote desktop client obsolete. This potentially increases usability while providing a largely platform-independent approach, as only a web browser is required on users' devices.

\subsection{Remote Laboratory Virtualization}
The laboratory hardware and development tools are accessed via virtual machines provided by an \emph{Oracle VM VirtualBox} hypervisor. This offers a range of advantages: Firstly, flexibility and scalability are increased. Different hardware configurations and corresponding software requirements can be implemented independently using different virtual machines, which supports a strict separation of concerns. If additional resources are required, these virtual machines can be duplicated, distributed and deployed independently of the host machines' architecture with minimal configuration effort. This also allows the distribution of pre-built images, provided this is permitted by the licences of the bundled software. Secondly, a higher level of security is being pursued. Users' influence is limited to the virtual machine assigned to them, which reduces attack vectors both on the host machine and the working environment of other users. In addition, the use of immutable disk images makes it possible to reset the machines to their original state after each session in order to guarantee a clean, functioning environment for subsequent users. This can be done automatically using a scheduled task, thus reducing administrative effort.

The system was designed in such a way that components are loosely coupled both in terms of the specific implementation (e.g. different hypervisors, operating systems or remote desktop servers can be used) and the site of deployment, providing further scalability prospects. The latter is facilitated by the use of a reverse proxy server, which orchestrates communication between clients and the various distributed components and services.

\section{Evaluation}
The system presented is being used for an initial evaluation as part of the course "Introduction to FPGA board development" at the University of Siegen. The web application represents an alternative to the existing system for the students. Both systems are therefore offered in parallel. 

Currently, 42 students have access to the new system and have organized themselves into 20 groups. Each group consists of two to five students. The system will enable students to collaborate on assignments covering topics such as input/output, memory architecture, and debugging using FPGA boards. For this purpose, a total of three \emph{Digilent Zybo} boards, virtual machines providing the \emph{AMD Xilinx Vivado Design Suite} development environment, an \emph{Arduino}-based sensor and relay assembly, and a webcam are made available to the students via the web application.

\section{Conclusion}
The architecture that has been implemented and is now ready for evaluation offers advantages for FPGA boards that go beyond the defined requirements \cite{alamin.2023}:
\begin{enumerate}
    \item Work sessions facilitate collaboration among students in groups in all aspects.
    \item Groups can be created in various sizes and the use of existing hardware can be maximized  by setting the number of bookable time slots per group per week.
    \item The infrastructure is highly scalable and flexible, utilizing virtual machines that can be distributed across the network.
\end{enumerate}

We aim to achieve significant breakthroughs in remote lab research by incorporating the collaborative features described above. Currently, we are also working on integrating our services for various use cases, including electrical engineering education.

\bibliographystyle{IEEEtran}
\bibliography{Design_and_Development_of_a_Multi-Purpose_Collaborative_Remote_Laboratory_Platform}

\end{document}